\documentclass[aip,pop,reprint]{revtex4-1}
\usepackage[utf8]{inputenc} % Выбор языка и кодировки
\usepackage[pdftex]{graphicx}
\usepackage{amsmath}
\usepackage{amssymb} % for \ggg
\usepackage{siunitx}

\graphicspath{{./figs/}}

\begin{document}

\title{Incoherent synchrotron emission of laser-driven plasma edge}

\date{\today}
\author{D.~A.~Serebryakov}
\email{dmserebr@gmail.com}
\author{E.~N.~Nerush}
\author{I.~Yu.~Kostyukov}
\affiliation{Institute of Applied Physics of the Russian Academy of
Sciences, 46 Ulyanov St., Nizhny Novgorod 603950, Russia}
\affiliation{Nizhny Novgorod State University, 23 Gagarin Avenue, Nizhny
Novgorod 603950, Russia}

\begin{abstract}

When a relativistically intense linearly polarized laser
pulse is incident on an overdense plasma, a dense electron layer is formed on the plasma edge which relativistic motion results in high harmonic generation, ion acceleration and incoherent synchrotron emission of gamma-photons. Here we present a self-consistent analytical model that describes the edge motion and apply it to the problem of incoherent synchrotron
emission by ultrarelativistic plasma electrons. The model takes into account both coherent radiation reaction from high harmonics and incoherent radiation reaction in the Landau-Lifshitz form. The analytical results
are in agreement with 3D particle-in-cell simulations in a certain parameter region that corresponds to the relativistic electronic spring interaction regime. 

%Electron heating and motion of plasma ions may smooth the
%plasma edge and affect electron dynamics that is not taken into account by the
%model

\end{abstract}

\maketitle

\section{Introduction}

Gamma-rays and hard X-rays have become widely applied since their
discovery. Most of present-day gamma-ray sources
are based on the radioactive decay, bremsstrahlung, and backward Compton
scattering; however, one needs to deal with radioactive materials
or large-scale electron accelerators (linacs or synchrotrons) to use them and it limits their avaliability.
The growing demand for gamma-ray sources in numerous fields (medicine,
radiography, nuclear physics) drives search for more convenient
and accessible sources of hard X-rays and gamma-rays. In the recent decades an
outstanding progress in laser technologies has been achieved, and now high-power 
(\SI{> 100}{TW}) short-pulse lasers have become commercially available. Due to ultrahigh 
electromagnetic fields that they produce, lasers may be used 
to improve brightness and flux of gamma-ray sources. Namely,
laser wake field acceleration (LWFA) is used to produce high-charge electron
bunches for bremsstrahlung ~\cite{Glinec05, Cipiccia12} and
Compton sources~\cite{Phuoc12, Gizzi13, Chen13a, Powers14,
Sarri14}; some other techniques are also proposed~\cite{Cipiccia11,
Chen13, Andriyash14}, one of them is incoherent synchrotron
emission of plasmas lit by high-power laser pulses.
%TODO: what about the other?

When a relativistically intense laser pulse interacts with a target,
the target electrons are expelled from atoms and accelerated to
relativistic speeds, then the laser field forces them to emit photons
due to Compton scattering. If the laser field amplitude $E_0$ is
such that $ a_0 \equiv e E_0 / m c \omega \ggg 1$ (where $c$ is the speed of
light, $\omega$ is the laser field angular frequency, $m$ and $e>0$ are the
electron mass and charge, respectively), the spectrum of emitted
photons becomes synchrotron-like with a tail up to \SI{}{\MeV} or above.
Earlier estimations and numerical simulations show that a large fraction of the laser pulse
energy may be transformed into gamma-rays if laser intensity is
high enough, e.g. percents~\cite{Nakamura12, Brady12, Nerush14} for $
I \gtrsim \SI{e22}{\W \thinspace \cm^{-2}}$ and tens of
percent~\cite{Ridgers12, Ji14a, Brady14a} for $ I \gtrsim \SI{e24}{\W
\thinspace \cm^{-2}}$, so such gamma-ray sources look
very promising (compared to linear Compton scattering-based sources with lower conversion efficiency,
 and to relatively complicated LWFA sources). But these phenomena are
commonly studied by particle-in-cell codes, and currently a few analytical models are
proposed.

In this paper we present a self-consistent analytical model describing nonlinear electron motion
in a thin dense layer arising at a laser-irradiated plasma edge in a specific
parameter region. The corresponding laser-plasma interaction regime is referred
as the {\it relativistic electronic spring}~\cite{Gonoskov11} (RES), and is mostly
considered in the context of attosecond pulse
generation~\cite{Gonoskov11, Brugge10}. In the RES regime a sizable portion of the
laser energy is transferred into electron oscillations~\cite{Gonoskov11} that
seems preferable for efficient photon emission. The thickness of the electron
layer in the RES regime may be as small as \SI{1}{\nm}, which makes coherent
synchrotron emission feasible~\cite{Gonoskov11, Brugge10, Dromey12}. %TODO: more understandable
The wavelength of \SI{1}{\nm} approximately corresponds to the photon energy of
\SI{1}{\keV}, therefore higher photon energies can be obtained only by
the {\it incoherent synchrotron emission} (ISE). However, for laser intensities $ I
\gtrsim \SI{e22}{\W \thinspace \cm^{-2} } $ average photon
energies of single electron synchrotron spectrum are far beyond the limit 
of coherent emission; that's why ISE can be very efficient.
We compute radiation pattern and other properties
of ISE in RES regime using the proposed analytical model for the plasma edge
dynamics.

Analytical results are verified by 3D particle-in-cell (PIC)
simulations that take into account photon emission with radiation reaction and
electron-positron pair production from hard photons~\cite{Nerush14}. Despite PIC
simulations reveal more complicated electron dynamics than assumed in the
model, the model describes well the plasma edge motion and the 
gamma-ray radiation pattern in a certain parameter region. Numerical
simulations also allow us to distinguish various laser-plasma interaction
regimes and compare gamma-ray generation efficiency between them.

\section{Analytical model formulation}

To describe collective electron dynamics and hard photon emission, let us 
introduce a self-consistent model of electron layer movement. Several major assumptions are pointed out. First, under the light pressure electrons of the irradiated plasma
edge form a very thin layer (in comparison with the laser wavelength) 
which moves in such a way that laser pulse reflects from the target almost completely, so we can 
neglect electron density perturbations behind the layer.
Second, we consider normal incidence of linearly polarized
laser pulses and assume that the layer electrons move in the polarization
plane. Third, only collective electron dynamics is considered (we suppose that the dispersion of individual electron characteristics inside the layer does not affect hard photon emission
drastically and just results in a smoothing of the radiation pattern, photon
spectrum, etc). The last assumption is that the ion motion is neglected: 
we restrict ourselves to the case of few-cycle laser pulses so that the
interaction time is short enough, or
to the case of quite low laser intensity so that the ion displacement is
negligible.

The total force driving the layer electrons consists of the following parts:
force caused by the incident laser field, force from self-generated
electromagnetic fields (i.e. fields coherently emitted by the layer), force
caused by electron-ion separation and radiation reaction force (i.e. force
caused by ISE). Unlike the model of Ref.~\onlinecite{Gonoskov11}, we do not adopt the
requirement that the laser field behind the layer is completely compensated by the
self-generated fields: this requirement makes it impossible to obtain Lorentz
factor of the electrons~\cite{Gonoskov11, Nerush14}. Instead, we start from the
equations of motion that takes into account all the force parts mentioned above:
\begin{eqnarray}
  \label{eq:model1}
  \frac{dp_x}{dt} = -E_x - v_y B_z + {F_r}_x, \\
  \frac{dp_y}{dt} = -E_y + v_x B_z + {F_r}_y, \\
  \mathbf E = \mathbf E_l + \mathbf E_p, \\
  \label{eq:model-1}
  \mathbf B = \mathbf B_l + \mathbf B_p,
\end{eqnarray}
where $\mathbf p$ is the electron momentum normalized to $mc$, $\mathbf v$ is
the electron velocity normalized to $c$, $\mathbf E$ and $\mathbf B$ are the
electric and magnetic fields, respectively, normalized to $mc\omega / e$, indexes $l$ and $p$ denote laser and
plasma fields (plasma fields are coherent emission and charge separation fields), and
$\mathbf F_r$ is the radiation reaction force (caused by ISE).
Here we assume that the laser pulse is incident along $x$ axis and the polarization
plane is the $xy$ plane.

The models similar to Eqs.~(\ref{eq:model1})-(\ref{eq:model-1})
have been described in Refs. \onlinecite{Rykovanov11,Bulanov13b}. They account only the radiation reaction
caused by the fields coherently emitted by the layer, i.e., transverse components of
$E_p$ and $B_p$. On the contrary, this model considers both the coherent part of radiation reaction (from coherent high harmonics emission)
and the radiation reaction force caused by the incoherent
emission of high-frequency photons. The wavelengths of these photons (emitted by
single electrons) are much smaller than the layer thickness so they cannot be
coherently summarized.

As stated above, we consider the incidence of a laser pulse on a plasma
half-space (unlike Ref.~\onlinecite{Bulanov13b} considering a foil much thinner
than the laser wavelength). Therefore, the areal charge density of the layer is
a function of the layer position, and according to RES
assumption~\cite{Gonoskov11} is the
following:
\begin{equation}
  \epsilon = n_0 x_\ell,
\end{equation}
where the charge density is normalized to $c n_{cr} / \omega$, $n_0$ is the
initial electron density normalized to the critical plasma density $n_{cr} = m
\omega^2 / 4 \pi e^2$, $x_\ell$ is the layer displacement from the initial
plasma edge position
normalized to $\lambda / 2 \pi = c / \omega$. The fields coherently emitted by
the layer can be then easily found from the Maxwell's equations~\cite{Brugge10,
Gonoskov11, Nerush14}:
\begin{eqnarray}
  E_{y,+} = B_{z,+} = \frac{\epsilon v_y}{2(1-v_x)},\\
  E_{y,-} = -B_{z,-} = \frac{\epsilon v_y}{2(1+v_x)},\\
  E_x = \frac{\epsilon x}{x_\ell} \text{ if } 0 < x < x_\ell,
\end{eqnarray}
where indices $+$ and $-$ denote field values at $x_\ell \pm 0$ and also denote
waves emitted by the layer in $+x$ and $-x$ directions. 
$E_x$ can be found as a flat capacitor field since the ions are immobile.
An example of the fields generated by the electron
layer oscillating according to Eqs.~(\ref{eq:xell1})-(\ref{eq:xell4}) (see
below) is shown in Fig.~\ref{fig:schematic}.

\begin{figure}
\includegraphics{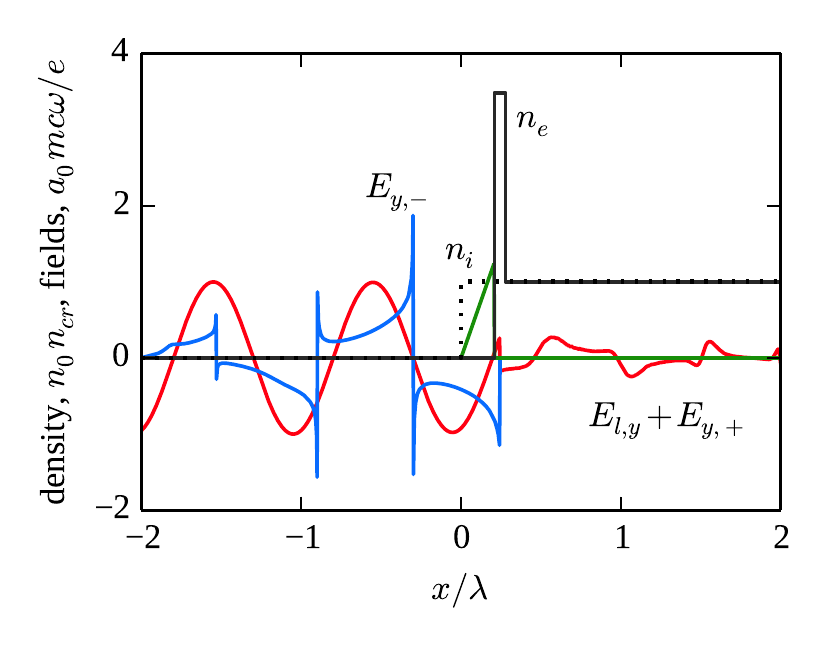}
\caption{\label{fig:schematic}Schematic structure of the electron layer at the
  plasma edge and fields in RES regime. Curves: electron density, gray; ion
  density, black dotted; $E_x$, green; sum of the laser field and the field
  emitted by the layer in $+x$ direction ($E_{l,y}+E_{y,+}$), red; the wave
  emitted by the layer in $-x$ direction ($E_{y,-}$), blue.}
\end{figure}

The radiation reaction force caused by ISE may become significant at field
intensities of $\gtrsim$\SIrange[range-units = single,
range-phrase=--]{e21}{e22}{\W/\cm^2} so it shouldn't be
neglected for the sake of consistency~\cite{Ridgers13, Tamburini12, Chen11, Capdessus12}. The main term
in Landau-Lifshitz approximation of the radiation reaction force is the
following~\cite{Landau75}:
\begin{equation}
  \label{eq:RFF} \mathbf{F}_r = -\frac{4 \pi}{3} \frac{r_e}{\lambda} \left[
  \left( \gamma \mathbf{E} + \mathbf{p} \times \mathbf{B} \right)^2 -
  \left(\mathbf{p} \cdot \mathbf{E} \right)^2 \right] \mathbf{v},
\end{equation}
where $r_e = e^2 / mc^2$ is the classical electron radius and
$\gamma = \sqrt{1 + p^2}$ is the electron Lorentz factor. The radiation
reaction efficiently decelerates the emitting electrons and
modifies their movement, however, we observe that the trajectory
doesn't change qualitatively. Anyway,
the Lorentz factor and its distribution along the electron trajectory is affected
by the radiation reaction that can yield modification of total power and radiation
pattern of the emitted gamma-rays. 

It can be easily estimated from Eq.~(\ref{eq:RFF}) that for a simple circular trajectory in rotating
electric field radiation reaction force is equal to the Lorentz force when
the dimensionless field magnitude reaches 
${a}_{thr} = \sqrt[3]{{3\lambda}/{4\pi r_e}} \approx 400$ for optical
wavelengths. For more complex trajectories radiation reaction
force can be higher than for circular trajectory. Anyway, for more strict
consideration radiation reaction force should be taken into account if
$a_0 \gtrsim 100-400$.

Since a field affecting the layer can be found as a half-sum of 
fields on both layer sides, the system of equations that governs 
collective electron dynamics turns to be
\begin{eqnarray}
\label{eq:xell1}
\frac{dp_x}{dt} = -\frac{n_0 x_\ell}{2}\left(1+\frac{v_x v_y^2}{1-v_x^2}\right)
- v_y {E_l}_y + {F_r}_x, \\
\label{eq:xell2}
\frac{dp_y}{dt} = -\frac{n_0 x_\ell v_y}{2} - ( 1 - v_x ) {E_l}_y + {F_r}_y, \\
\label{eq:xell3}
\frac{dx_\ell}{dt} = v_x = \frac{p_x}{\gamma_\ell}, \\
\label{eq:xell4}
\frac{dy_\ell}{dt} = v_y = \frac{p_y}{\gamma_\ell}.
\end{eqnarray}

\begin{figure}
\includegraphics{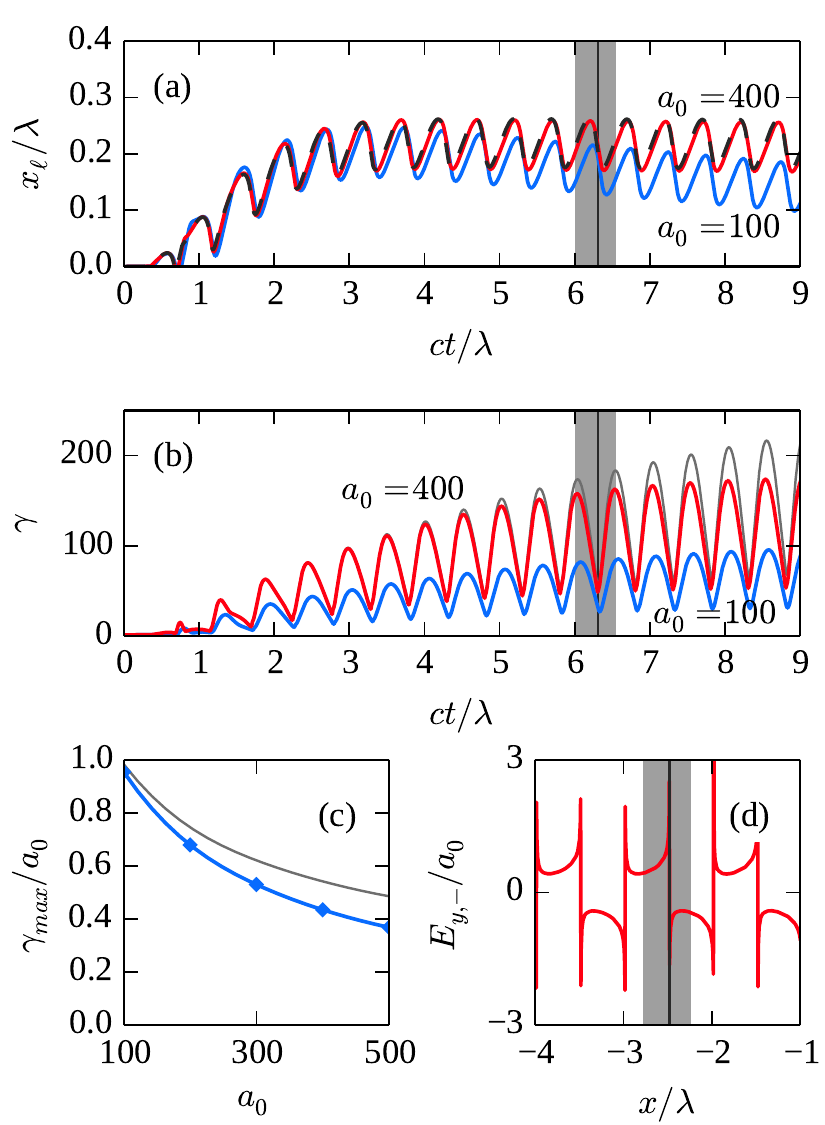}
\caption{\label{fig:gamma}(a) The layer position given by the model for $a_0 =
  n_0 = 400$ (red curve) and $a_0 = n_0 = 100$ (blue curve). The dashed black
  curve corresponds to the layer motion that yields a complete compensation of
  the incident field behind the layer~\cite{Nerush14} for $n_0 = a_0$. (b) The
  Lorentz factor of the above-mentioned trajectories. Thin grey curve
  corresponds to Eqs.~(\ref{eq:xell1})-(\ref{eq:xell4}) with $\mathbf F_r$
  neglected, $a_0 = n_0 = 400$. (c) Maximal Lorentz factor gained by the layer
  electrons for $a_0 = n_0$ at $9$ periods of the incident field vs the incident
  field magnitude according to the analytical model (blue curve with markers),
  and the same with $\mathbf F_r$ neglected (thin grey curve). (d) The field
  reflected by the layer, $a_0 = n_0 = 400$, at $t = 9 \lambda / c$.  The gray
  area denotes a trajectory half period bounded by $v_x \approx 1$ points,
  vertical dark grey line denotes a time instant where $v_x \approx -1$.
}
\end{figure}

\subsection{Some model properties}
\label{sec:modelProperties}

\begin{figure*}
\includegraphics{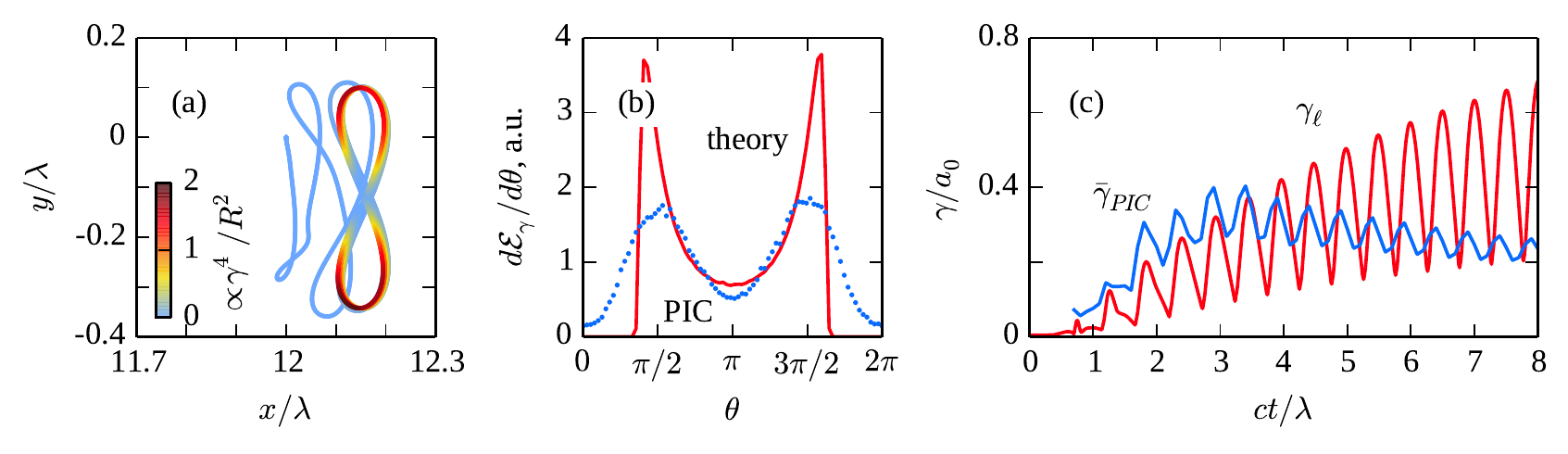}
\caption{\label{fig:nxt1}(a) The layer trajectory for $a_0 = 240$ and $n_0 =
  320$ obtained from Eqs.~(\ref{eq:xell1})-(\ref{eq:xell4}), color corresponds
  to $\gamma^4 / R^2$. (b) Radiation pattern for the theory (solid) and PIC
  (dotted); $\vartheta$ is the angle in the polarization plane, $\vartheta = 0$
  corresponds to the direction of $x$ axis. (c) Lorentz factor in the model
  $\gamma_\ell$, and the average Lorentz factor $\bar \gamma_{PIC}$ of electrons with 
  $\gamma > 0.05 a_0$.
}
\end{figure*}

In this subsection we neglect radiation reaction force $\mathbf F_r$ in order to
demonstrate clearly a number of results of the model. First, the equation
for Lorentz factor can be found from Eqs.~(\ref{eq:xell1})-(\ref{eq:xell4}) and
is the following:
\begin{equation}
  \label{eq:dgammadt}
  \frac{d\gamma_\ell}{dt} = -({E_l}_y + E_{y,+}) v_y - \frac{ n_0 x_\ell v_x }{ 2
  (1 + v_y^2 \gamma_\ell^2) }.
\end{equation}
Thus, the electron Lorentz factor is determined by the field transmitted
through the layer, ${E_l}_y + E_{y,+}$, and by the effective longitudinal field
reduced by a factor of $1 + v_y^2 \gamma_\ell^2$. Lets assume that $\gamma_\ell \gg
1$ and $|{E_l}_y + E_{y,+}| \ll a_0$, hence $d\gamma / dt \ll dp_x / dt$. In
this case Eqs.~(\ref{eq:xell1}) and (\ref{eq:xell2}) together with $d(v_x^2 +
v_y^2)/dt \approx 0$ yield: $v_y {E_l}_y + n_0 x_\ell (1 + v_x) / 2 \approx 0$. This
equation is equivalent to the following: ${E_l}_y + E_{y,+} \approx 0$.

Therefore, Eqs.~(\ref{eq:xell1})-(\ref{eq:xell4}) lead to the layer dynamics
that almost completely compensates the incident field behind itself. This means that the
trajectory obtained from the model is close to the layer trajectory obtained in
the framework of Refs.~\onlinecite{Gonoskov11, Nerush14} in which the layer
trajectory is found from the strict requirement ${E_l}_y + E_{y,+} = 0$, that
can be rewritten as follows~\cite{Nerush14}:
\begin{equation}
  \frac{d x_\ell}{dt} = \frac{4 {E_l}_y^2 - n_0 x_\ell^2}{4 {E_l}_y^2 + n_0
  x_\ell^2}.
\end{equation}
However, the latter model assumes that Lorentz-factor is constant
along the layer trajectory and doesn't allow to find the radiation pattern
since it highly depends on $\gamma_l$ distribution along the trajectory.

Since the terms of Eq.~(\ref{eq:dgammadt}) may be of the same order and the last
term should not be neglected, $a_0$ cannot be excluded from
Eqs.~(\ref{eq:xell1})-(\ref{eq:xell4}) by a normalization, and not the only parameter $n_0/a_0$
governs the layer dynamics, but both $a_0$ and $n_0$. This means that the scaling law proposed in
Refs.~\onlinecite{Gordienko05, Gonoskov11, Nerush14} can be invalid for
laser-plasma interactions in RES regime.

The described model properties are illustrated in Fig.~\ref{fig:gamma} where
numerical solution of Eqs.~(\ref{eq:xell1})-(\ref{eq:xell4}) is shown. First, it
is seen that $x_\ell$ given by the model of the present paper is close to the
layer position given by models of Refs.~\onlinecite{Gonoskov11, Nerush14}. Second, due
to the field compensation $\gamma_\ell$ increases with time quite slowly and
even for several laser periods $\gamma_\ell / a_0$ remains $\lesssim 1$.
Furthermore, $\gamma_\ell / a_0$ drops with the instantaneous increase of $a_0$
and $n_0$. Third, radiation reaction doesn't change the interaction picture
considerably.

\subsection{ISE in the model}

Eqs.~(\ref{eq:xell1})-(\ref{eq:xell4}) don't allow analytical solution and should be solved
numerically. Solution depends on the parameters $a_0$, $n_0$ and laser
pulse shape. 

\begin{figure}
\includegraphics{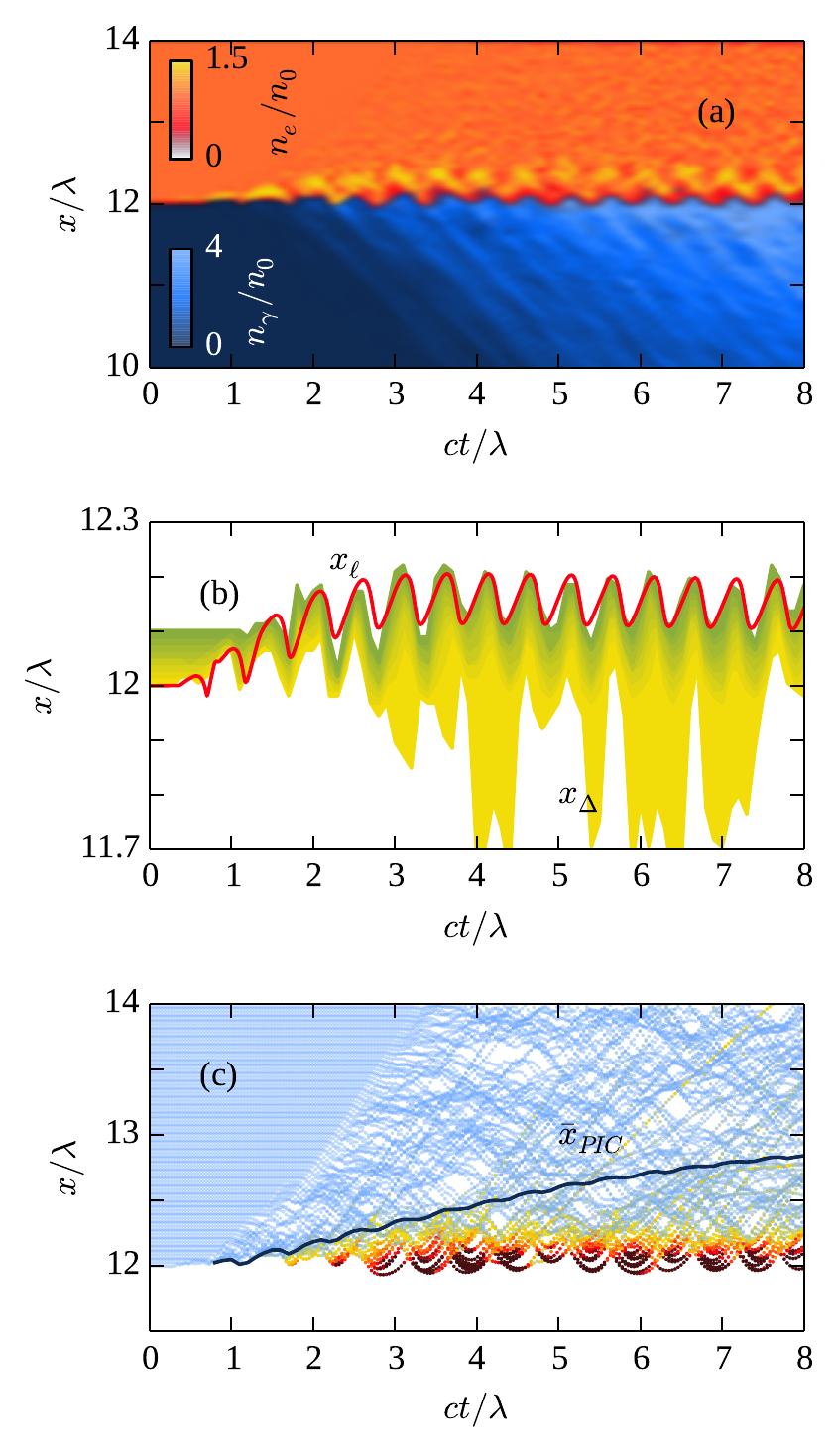}
\caption{\label{fig:nxt2}(a) On-axis electron and photon densities obtained in
  PIC for $a_0 = 240$ and $n_0 = 320$. (b) For the same parameters, the
  analytical trajectory (red) and $x_\Delta$ according to Eq.~(\ref{eq:xDelta})
  for $\Delta = 0.01 \lambda$ (lower boundary) and $\Delta = 0.11 \lambda$
  (upper boundary). (c) Tracks of individual electrons obtained in PIC
  simulation. The color intensity corresponds to $\gamma^4/R^2$. The average
  trajectory of the electrons with $\gamma > 0.05 a_0$, $\bar x_{PIC}$.
}
\end{figure}

The sample electron layer trajectory for typical modeling parameters is shown in Fig.~\ref{fig:nxt1}(a). The asymptotic trajectory resembles 
number $8$ and is similar to a particle trajectory in a linear-polarized EM wave (for a free particle it looks like number $8$ only in 
a proper reference frame, though). For $a_0=240$ the maximum value of Lorentz factor is 120 and corresponds to the middle part of 
trajectory (where curve intersects itself); the minimum $\gamma$ value is in the order of ten. The distribution of the layer energy over 
the trajectory yields the electron layer radiation pattern. Given that energy is ultrarelativistic over the whole period ($\gamma \gg 1$), 
radiation mechanism is almost purely synchrotron --- practically all gamma quanta are emitted in the tangential direction. 
So we can use the following formula for the emission power of a single electron at each instant of time:
\begin{equation}
	I_e = \frac{2 e^2 c \gamma^4}{3R^2}
\end{equation}
where $R$ is the curvature radius of the trajectory at the current point. For gamma quanta emitted by the whole layer we get
\begin{equation}
	\label{eq:syncrotron2}
	I_{layer} \sim\frac{\gamma^4}{R^2} X_l
\end{equation}
On-axis distribution of the hard photon density is shown in Fig.~\ref{fig:nxt2}(a), and the radiation pattern is shown in Fig.~\ref{fig:nxt1}(b). 
On the curved part of the trajectory the gamma-rays are generated most efficiently due to small curvature radius, but the radiation direction 
changes rapidly at this point. It's the central part of the trajectory that mostly contributes into the radiation pattern.

The gamma ray radiation pattern shown in Fig~\ref{fig:nxt1}(b) corresponds to
the trajectory from Fig.~\ref{fig:nxt1}(a). The dotted line is the pattern that
was produced by PIC 3D numerical simulations (see Sec.~\ref{sec:pic} for
details). The pattern has two lobes in the plane of the laser $\mathbf{E}$ and $\mathbf{k}$ vectors. Lobes position depends on the laser pulse intensity;
with greater intensity, the maximums are closer to the $x$-direction. In this
model, nothing is emitted strictly forward because the layer electrons are
 never moving exactly in the $+x$ direction. Qualitatively the model and numerical results are
in agreement with each other, however simulations don't give sharp peaks in the radiation pattern due to electron velocity spread which is neglected in model.

\section{Results of numerical simulations} \label{sec:pic}

In order to validate the model results, we have performed a series of numerical PIC simulations. The code used for is fully three-dimensional and 
can take into account the quantum electrodynamical (QED) effects~\cite{Nerush14}. The code supports parallelization so it can be efficiently used on computational clusters.

We have simulated the interaction of a falling \SI{1}{\um} wavelength linear-polarized laser pulse with a flat \SI{2}{\um} wavelength foil. 
The laser pulse intensity and foil density varied between different PIC runs. In this series, we considered the ions as fixed; conditions of applicability of this assumption are discussed in the section \ref{sec:conditions}.
Laser pulse envelope has the following shape:
\begin{equation}
E_y(x)=\frac{d}{dx}\left\lbrace\sin{x}\cos^2\left(\frac{\pi(x+x_s)^4}{2x_s^4}\right)\right\rbrace
\end{equation}
It corresponds to a wavepacket which has almost constant amplitude in the central area and promptly decreases at the $x_s$ distance from the pulse center. Therefore the configuration becomes much close to idealized modeling situation (falling of a plain wave onto a flat target).

\subsection{Electron layer dynamics}
In a certain region of simulation parameters a layer with high electron density is formed in the front side of the target 
(see Sec.~\ref{sec:conditions} for details). Its thickness is in the order of \SI{100}{\nm}. Due to $\mathbf v
\times \mathbf B$ force caused by linearly polarized laser pulse the plasma edge
is oscillating in the direction of pulse propagation (see
Fig.~\ref{fig:nxt2}(a)). Electrostatic force caused by the ions compensates the light
pressure on the average and holds the layer.

From the numerical simulations we can acquire the layer trajectory and compare it to the
analytical one. Current position $x_\Delta$ of the layer in PIC is calculated using the following formula
\begin{equation}
\label{eq:xDelta}
\Delta = \int_{-\infty}^{x_\Delta} n_e(x) / n_{cr} \, dx,
\end{equation}
where $\Delta$ is some given in advance number of electrons. If the electrons form a
thin layer and $\Delta$ is less than a number of electrons in the layer, 
$x_\Delta$ is approximately equal to the layer position for any value of
$\Delta$. It is seen in Fig.~\ref{fig:nxt2}(b) that for $0.05 \lesssim
\Delta/\lambda \lesssim 0.1$ (shown by green), $x_\Delta$ obtained from
Eq.~(\ref{eq:xDelta}) is in good agreement with
the analytical model. It is also seen that a small number of
electrons can be pulled out from the plasma (see $\Delta \lesssim 0.05 \lambda$,
shown in yellow, in Fig.~\ref{fig:nxt2}(b)).

However, the layer dynamics in PIC simulations doesn't fully obey theoretical assumptions. From Fig.~\ref{fig:nxt2}(c) we can see 
that the layer is an average characteristic because single electrons constantly join and leave the layer. A typical electron
lives in the layer one-two laser field periods and then enters the quasi-unperturbed plasma; it results in decrease of
the layer energy compared to the theory (see $\bar \gamma_{PIC}$ on Fig.~\ref{fig:nxt1}(c)) and modification of spatial distribution of energetic
electrons (see $\bar x_{PIC}$ on Fig.~\ref{fig:nxt2}(c)). It alters the energetic properties of electrons and gamma-quanta in PIC compared to the theory.

\subsection{Energetic properties and scaling laws}

\begin{figure}
	\includegraphics[width=\columnwidth]{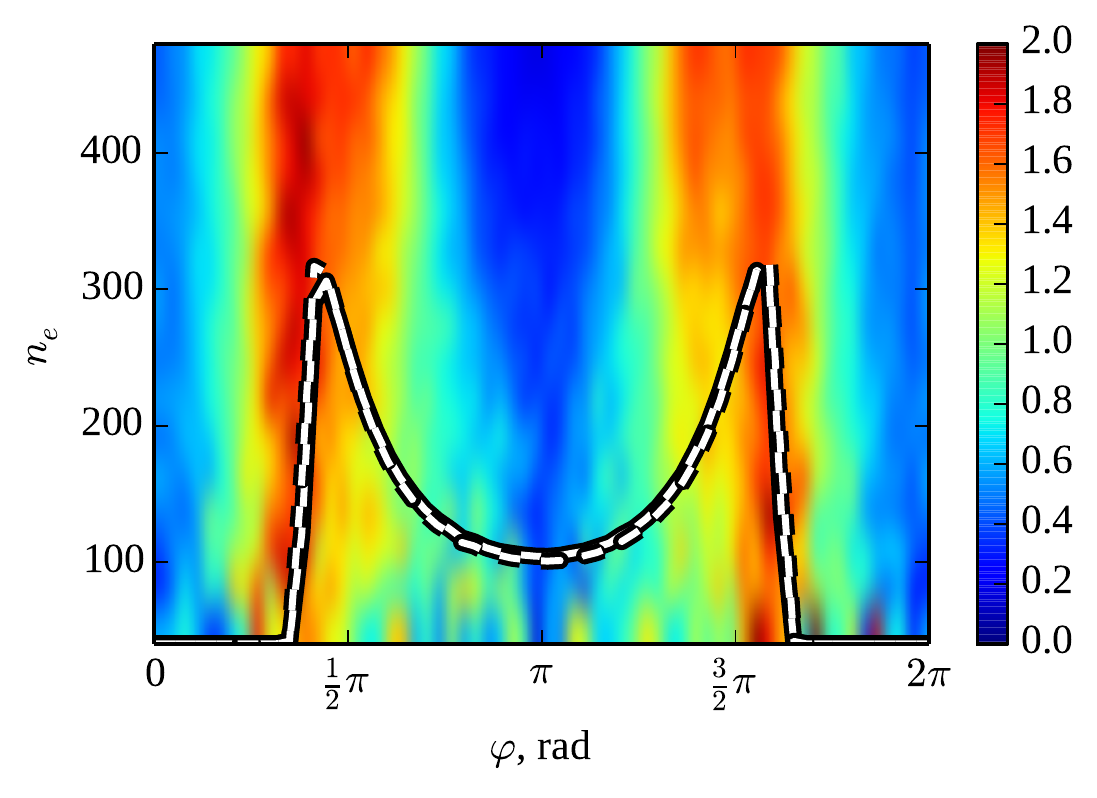}
	\caption{\label{fig:rp-fromS}Gamma-ray radiation pattern in PIC simulations at $a_0=n_0$ ($S=1$) varying from 40 to 480; 
	lines show radiation pattern from the model for $S=1$, $a_0=120$ (solid line), $a_0=240$ (dashed).}
\end{figure}

One of well-known features of many laser-plasma problems is that at relativistic
laser intensities there is a similarity law \cite{Gordienko05}: the interaction
properties depend on the dimensionless parameter 
\begin{equation}
S = \frac{n_e}{a_0 n_{cr}} = \frac{n_0}{a_0}
\end{equation}
and not on $n_0$ and $a_0$ separately. The current model doesn't obey this
scaling law strictly: equations (\ref{eq:xell1}-\ref{eq:xell4}) can't be reduced
to a form with one dimensionless parameter even if radiation reaction is
neglected (see also Sec.~\ref{sec:modelProperties}). However, the dependence on
$a_0$ at $S=\operatorname {const}$ is slight. We have compared radiation patterns from model and PIC simulations at different 
laser pulse intensities and foil densities but the same $S$ parameters and the difference is negligible (see Fig. \ref{fig:rp-fromS}).

\begin{figure}
\includegraphics[width=0.5\textwidth]{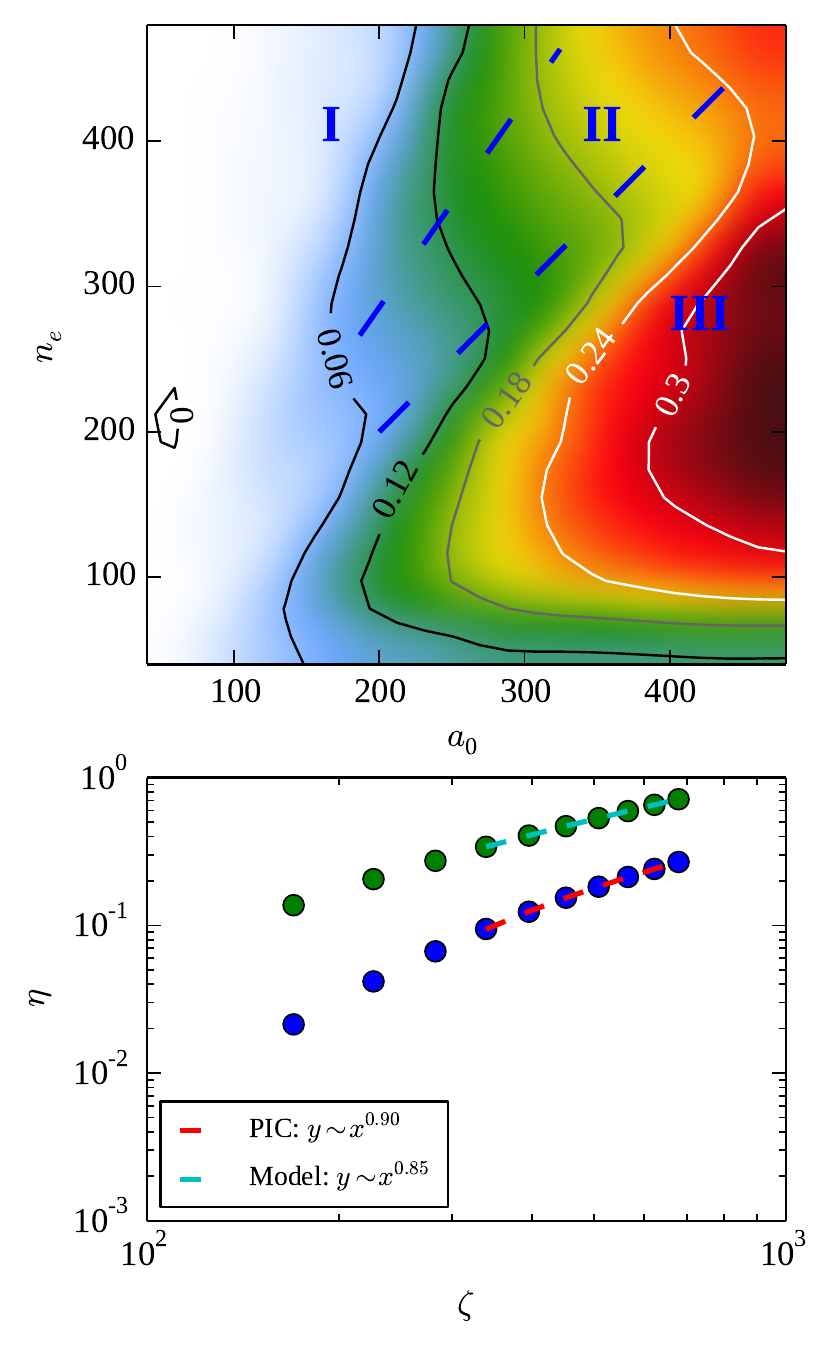}
\caption{\label{fig:mathcalE} (a) Gamma-ray generation efficiency map obtained from 144 PIC simulations. Foil thickness is \SI{2}{\um}, foil density and laser pulse intensity are variable parameters. II is the region where the model is applicable;  III is the region of relativistically self-induced transparency (RSIT), in region I the stable electron layer is not formed in PIC.
(b) Gamma-ray generation efficiency from PIC and analytical model at $S=1$ ($a_0=n_0$).}

\includegraphics{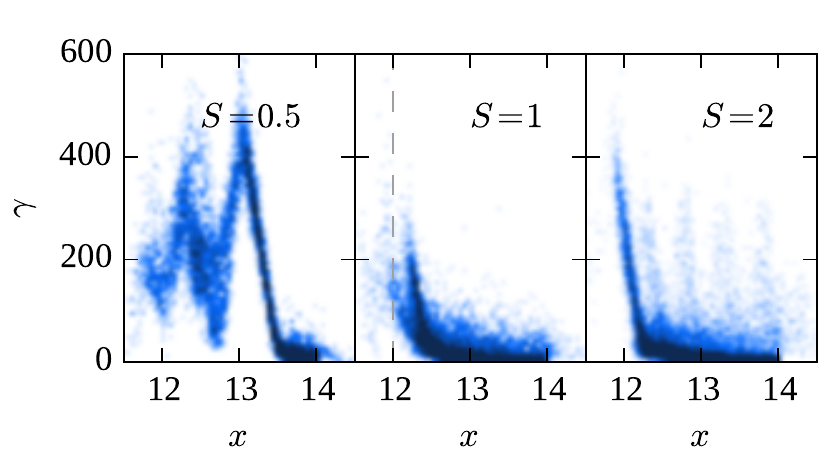}
\caption{\label{fig:xg-regimes} Electron Lorentz-factor distribution along
  $x$-axis at $t = 5.8 \lambda / c$ (laser pulse is incident from the left) for
  different values of $S$. Only electrons from the region near the $x$ axis,
  where laser pulse front is plane, are shown.
}
\end{figure}

Figure \ref{fig:mathcalE}(a) shows gamma-ray generation efficiency map in the $n_0-a_0$ axis which is acquired from 144 PIC simulations; 
each point corresponds to the energy of gamma-rays generated during simulation timespan, normalized to the energy of the laser pulse. 
Different values of similarity parameter $S$ correspond to sloping lines starting at the coordinate origin, it can be clearly seen from 
the figure that $S$ determines different regimes of gamma-ray generation. 

Higher $a_0$ values always result in increase of gamma-ray generation efficiency. The parameter region $S<1$ (which is denoted as "I") 
corresponds to the relativistically self-induced transparency (RSIT) regime; the target becomes transparent because of the relativistic 
electron mass reduction and effective $a_0$-times decrease of plasma frequency and increase of plasma density the electromagnetic waves 
can propagate through~\cite{Kaw70}. Penetrating laser pulse produces extreme heating of electrons and ions in plasma and causes emission 
of hard photons due to ISE. In this region gamma-rays are generated most efficiently; however, the model is not applicable here because of 
RSIT and absence of the single electron layer being considered (we will discuss applicability conditions in more detail in the Sec. \ref{sec:conditions}).

For $S \ge 1$, we can compare generation efficiency from PIC and model, the example of comparison under constant $S=1$ and variable $a_0$ 
is shown in the \ref{fig:mathcalE}(b). From the model, the efficiency is about one order less that from PIC simulations. However, the scaling 
corresponds to a power law of the approximately same degree in both cases that turns to be an important result. Possible reasons for that: 
first, electrons in PIC simulations are not monoenergetic and have a certain Lorentz-factor distribution, so there is always a number of 
electrons that are hotter than the layer electrons in the model. Due to the radiation law $W \sim \gamma^4$, their contribution into the 
total energy output can essentially increase the gamma-ray generation power. Second, the model estimation for the number of electrons in 
the layer $n_e = n_0 x_\ell$ is obviously not true when layer disposition $x_\ell$ is about zero or even negative. In PIC simulations, 
the layer electrons sometimes had near-zero disposition, so model underestimates radiation of electrons in this part of trajectory.

\section{Conditions of applicability} \label{sec:conditions}

%TODO: check layer form in PIC again
Based on 3D PIC simulations, we can distinguish a parameter region where the
developed model can be applied. The main assumption of the model is that the
incident laser pulse pushes the plasma electrons which form a single thin layer that reflects the laser pulse well so electrons behind the layer are
almost unperturbed. We perform the analysis of electron Lorentz-factor
distribution along longitudinal axis in different regimes
(Fig.~\ref{fig:xg-regimes}). It can be seen that the case of $S \sim 1$
corresponds to the model best. For $S<1$, RSIT leads to effective propagation of
the laser pulse through the target, and several electronic structures with high
values of Lorentz-factor are formed across the target. For $S \gtrsim 2$,
numerical simulations show that electron dynamics significantly differs from the
case of $S\sim1$ (where individual electrons stay in the layer up to several
laser periods): the electrons generally escape the layer on each of laser
half-periods, and new (`cold') electrons from the plasma replace them. Therefore
electrons are grouped in thin bunches which propagate in $x$-direction (see
Fig.~\ref{fig:xg-regimes},~(c)). This effect leads to effective increase of
overall electrons energy and decrease of gamma-rays generation efficiency
because electrons in the layer do not reach high energies during laser
half-period.

Therefore we define the region where the model can be applied as $1\le S <2$.
Limitation on $a_0$ parameter can be found from the requirement that we neglect
quantum electrodynamical effects in this model so $a_0$ shouldn't be greater
than $\sim 500$. Obviously the laser pulse should be also strong enough so that
the electron layer with high Lorentz-factor can be formed at all (thus $a_0 \gg
1$).

One should consider attainable plasma densities as well. For solid targets a typical electron density is in the order of 
$10^{22}-10^{23} \SI{}{\cm}^{-3}$ or higher so $n_0$ is in the order of hundreds
(if $\lambda ~ \SI{1}{\um}$). In order to achieve lower densities, some exotic targets are required.

The model also has the requirement of immobile ions. The numerical simulations that have been presented above are carried with immobile 
ions too, but several numerical experiments with moving ions were done.
 Charge-to-mass ratio of the ions was 0.25 of that 
for hydrogen atoms. The results (gamma-ray radiation pattern, electron layer dynamics) didn't change drastically in case of moving ions, 
although gamma-ray generation efficiency became lower about 2 times. To get mostly aligned with the model and 
improve gamma-ray generation efficiency, one should take as smallest ion charge-to-mass ratio as possible.

\section{Conclusion}

The self-consistent model that describes electron dynamics in the interaction of
a relativistically strong laser field with overdense plasma halfspace is presented.
Plasma electrons are pushed by $\mathbf v \times \mathbf B$ force and form a
thin layer on the plasma edge. This layer coherently emit electromagnetic spikes
as the laser pulse reflection and considerably suppress the fields transmitted
into the plasma. The back-reaction of the coherently emitted fields is taken
into account, as well as radiation reaction caused by the incoherent emission of
the layer.

The back-reflected field presents a train of electromagnetic spikes emitted when
the electron layer has maximal velocity towards the laser pulse. These time
instances correspond to the minimal value of the electron Lorentz factor that
means not optimal conditions of short-spike generation and low value of the
cut-off in the spectrum of the reflected field~\cite{Brugge10, Gonoskov11}.

When electrons are ultrarelativistic, the supposition that the plasma dynamics
is governed by the only parameter $S = n_0 / a_0$ (not by $n_0$ and $a_0$
separately) is often used~\cite{Gordienko05, Gonoskov11, Nerush14}. It
yields the electron Lorentz factor $\gamma \propto a_0$. But the
presented theory demonstrates that the process of electron acceleration can be a
fine effect. Namely, it occurs that $d\gamma/dt = 0$ if we neglect the electron rest
mass and fields behind the layer in the framework of the presented
theory~\cite{Nerush14}. The
accurate analysis shows that the equation for the electron energy depends on two
parameters $n_0$ and $a_0$ separately, and yields that the ratio $\gamma / a_0$
drops with the increase of $a_0$. The losses caused by the incoherent photon
emission strengthen this drop.

The presented model allows one to find electron trajectories (including Lorentz
factors) and, hence, incoherent synchrotron emission from the plasma edge.
Plasma edge dynamics and gamma-ray radiation pattern are in a fairly good
agreement with the results of 3D PIC simulations. Namely, the model predicts a
two-lobe radiation pattern with lobes approximately parallel to the electric
field direction of the incident laser pulse. This radiation pattern is caused by
the 8-like electron trajectories extended along this direction. It is also
predicted that for $S \approx 1$, in contrast to the electron reinjection
regime~\cite{Brady12}, emission towards the incident laser pulse is much weaker than
in perpendicular direction, that is also in good agreement with 3D PIC simulations. Therefore,
angular distribution of gamma-rays can specify the laser-plasma interaction
regime in experiments at ultrahigh intensity.

PIC simulations reveal that a thin electron layer is formed on the plasma edge
if $S = n_0 / a_0 \approx 1$. In the case of $S \lesssim 1/2$ relativistically self-induced transparency occurs, and the laser pulse
propagates through the plasma. If $S \gtrsim 2$, the skin depth (the thickness of
the current distribution in the plasma) becomes larger than electron
displacement perpendicular to the plasma surface, and a lot of electrons escapes
to the plasma bulk. If the laser pulse duration or the laser pulse magnitude $a_0$ is
high, ion motion becomes considerable. This case is also beyond the region where the model is applicable.

From numerical simulations we also see that individual
electrons stay near the plasma edge generally a half of the laser period, thus,
electrons permanently leave and join the layer. Electrons that leave the layer
transfer the laser energy into the plasma bulk and heat the target. Electrons
that join the layer have energy which is significantly different from the average electron
energy in the layer that may cause a sizable discrepancy between model
predictions and PIC results. Namely, PIC and the model gives different values
for overall electron energy and generation efficiency of gamma rays.
Nevertheless, the analytical model yields the same scaling of the generation
efficiency as PIC if $S \approx 1$.

Therefore, the presented theoretical model can be used for the analysis of
coherent and incoherent photon emission for $n_0 \sim a_0$. The model can be
used for optimization of high harmonic and gamma-ray generation in the interaction
of ultraintense laser pulses with plasmas. Some results of this work would be
extended to the case of obliquely incident laser pulses.

\section{Acknowledgements}

Numerical simulations were carried out on the computer cluster of the Laboratory of
Supercomputer Technologies in Non-Linear Optics, Plasma Physics and
Astrophysics in University of Nizhny Novgorod.

This work has been supported by the Government of the Russian Federation (Project No. 14.B25.31.0008)
and by the Russian Foundation for Basic Research (Grant No. 15-02-06079).

% TODO: add \bar \gamma_{PIC} description
% gamma in the layer from PIC - in Fig. 3(c)?
% plot photon spectrum!
% Ref to Luo15
% Ref to GonoskovPRL14, ref to GonoskovArXiv14
%
%radiation reaction force, which
%should be considered at relatively high laser field intensities
%(dimensionless field amplitude is in the order of hundreds)
%
% ref to Bulanov13a
%
% self-consistent
%
% Ref. to Shen01 (Meyer-ter-Vehn)
% Ref. to Pan15 (oblique incidense)
% Ref. to Capdessus13, 14.
% Fig. 5: on-axis electrons only, cmap instead of scatter, t=?, check phrase
% about skin-layer in the Conclusion
%Density of generated gamma-quanta is shown with blue in Fig.~\ref{fig:nxt}(b).?

\bibliography{article2014}
\end{document}